\documentclass[english,12pt,a4paper]{article}
\usepackage{authblk}
\usepackage[text={17.0cm,23.7cm}]{geometry}
\usepackage[T1]{fontenc}
\usepackage[utf8]{inputenc}
\usepackage{babel}
\usepackage{hyperref}
\usepackage{cite}
\usepackage{subfigure}
\usepackage{graphicx}
\usepackage{amsmath}
\usepackage{amssymb}
\usepackage[dvipsnames]{xcolor}
\usepackage{yfonts}
\usepackage{ulem}
\usepackage{enumitem}
\usepackage[mathscr]{euscript}
\usepackage{float}
\usepackage{physics}
\usepackage{slashed}

\begin{document}
	\title{\vspace{-2cm}
		{\normalsize
			\flushright TUM-HEP 1355/21\\}
		\vspace{0.6cm}
		\textbf{Flavor violating muon decay into an electron and a light gauge boson}\\[8mm]}
	\author[1]{Alejandro Ibarra}
	\author[2]{Marcela Mar\'in}
	\author[2]{Pablo Roig}
	\affil[1]{\normalsize\textit{Physik-Department, Technische Universit\"at M\"unchen, James-Franck-Stra\ss{}e, 85748 Garching, Germany}}
	\affil[2]{\normalsize\textit{Centro de Investigaci\'on y de Estudios Avanzados del Instituto Politécnico Nacional, Apartado Postal 14-740, 07000, Ciudad de M\'exico, M\'exico.}}
	
	\date{}
	
	\maketitle
	\begin{abstract}
	We analyze the flavor violating muon decay $\mu\rightarrow e\chi$, where $\chi$ is a massive gauge boson, with emphasis in the regime where $\chi$ is ultralight. We first study this process from an effective field theory standpoint in terms of form factors. We then present two explicit models where $\mu\rightarrow e\chi$ is generated at tree level and at the one-loop level. We also comment on the prospects of observing the process $\mu\rightarrow e\chi$ in view of the current limits on $\mu\rightarrow 3e$ from the SINDRUM collaboration. 
	\end{abstract}

\section{Introduction}

The Standard Model of Particle Physics \cite{Glashow:1961tr, Weinberg:1967tq, Salam:1968rm} predicts the conservation of lepton flavor. The discovery of neutrino oscillations \cite{Fukuda:1998mi, Ahmad:2001an, Ahmad:2002jz} provided conclusive evidence for the violation of lepton flavor in Nature, and therefore for the existence of new physics beyond the Standard Model. However, and aside for neutrino oscillations, no other lepton flavor violating process has been observed up to this day. 

Experimental searches for lepton flavor violation in muon decays date back to the late 1940's \cite{Hincks:1948vr,PhysRev.74.1364}. The most sensitive searches as of today have set the upper limits ${\rm Br}(\mu^+\rightarrow e^+\gamma)\leq 4.2\times 10^{-13}$ by the MEG collaboration~\cite{TheMEG:2016wtm} and  ${\rm Br}(\mu^+\rightarrow e^+e^- e^+)\leq 1.0\times 10^{-12}$ by the SINDRUM collaboration~\cite{Bellgardt:1987du}. The searches for rare muon decays are complemented by those for other lepton flavor violating processes, such as $\mu-e$ conversion in nuclei~\cite{Honecker:1996zf,Kaulard:1998rb,Bertl:2006up}, muonium-antimuonium conversion~\cite{Willmann:1998gd}, or in neutral meson decays, such as $K_L^0\rightarrow \mu^\pm e^\mp$~\cite{Ambrose:1998us} or $B^0\rightarrow \mu^\pm e^\mp$~\cite{Aaij:2013cby} among others (see {\it e.g.} \cite{Calibbi:2017uvl}).

In recent years there has been interest in Physics at the low energy frontier (for a review, see \cite{Jaeckel:2010ni}). This possibility may lead to new lepton flavor violating muon decays, such as $\mu\rightarrow e \chi$, with $\chi$ an invisible boson. The non-observation of this decay by the TWIST collaboration allows to set the 90\% C.L. upper limit $BR(\mu\to e\chi)< 8.1 \times 10^{-6}$ \cite{Bayes:2014lxz}. Further, the light boson $\chi$ could lead to the three-body lepton flavor violating decay  $\mu^+\rightarrow e^+e^- e^+$, when $\chi$ is off-shell, resulting into complementary constraints on this scenario. 

In this work we will focus on the possibility that $\chi$ is a light gauge boson, associated to the spontaneous breaking of an Abelian gauge symmetry, $U(1)_\chi$ (the case where $\chi$ is a light scalar or a pseudoscalar has been extensively studied, see {\it e.g.} \cite{Wilczek:1982rv, Grinstein:1985rt, Berezhiani:1989fp, Feng:1997tn, Hirsch:2009ee, Jaeckel:2013uva, Celis:2014iua, Celis:2014jua, Galon:2016bka, Calibbi:2016hwq, Ema:2016ops, Bjorkeroth:2018dzu, Bauer:2019gfk, Heeck:2019guh,  Cornella:2019uxs,Calibbi2021}).

The simplest Lagrangian describing the lepton flavor interacting interaction is ${\cal L}_{\rm LFV}= g \overline \mu \gamma^\rho \chi_\rho e +{\rm h.c.}$, with $\chi_\rho$ the 4-potential associated to the $U(1)_\chi$ symmetry. With this effective description, the rate for $\mu\rightarrow e\chi$ contains terms proportional to $g^2/m_\chi^2$, due to the emission of the longitudinal component of the gauge boson (a similar behavior is found for other effective interactions). Naively, the rate diverges as $m_\chi\rightarrow 0$, therefore the effective theory cannot be matched to the well studied decay $\mu\rightarrow e\gamma$. Further, one may wonder whether the decays into several gauge bosons $\mu\rightarrow e \chi\cdots\chi$, could also contribute significantly to the total decay width, reminiscent of the ``hyperphoton catastrophe'' for the electron decay into a neutrino and an ultralight photon~\cite{Okun:1978xn, Okun:1978bp, Suzuki:1988bd}. 

In this paper we present a detailed analysis of the decay rate $\mu\rightarrow e\chi$, with special emphasis in the regime where $m_\chi\rightarrow 0$ (for previous works, see {\it e.g.} \cite{Farzan:2015hkd,Heeck:2016xkh,Farzan:2016wym}). In Section \ref{sec:eff_theory} we present the most general effective interaction leading to the decay $\mu\rightarrow e\chi$ in terms of a number of form factors, and we identify the conditions that the form factors must fulfill in order to render a finite rate in the limit $m_\chi\rightarrow 0$. In Sections \ref{sec:tree-level} and \ref{sec:one-loop} we present two gauge invariant and renormalizable models where the process $\mu\rightarrow e\chi$ is generated, respectively, at tree level and at the one-loop level. We calculate the rate for $\mu\rightarrow e\chi$ and we explicitly show that the rate remains finite as $m_\chi\rightarrow 0$. For these two models, we also calculate the rate for the rare decay $\mu\rightarrow 3e$. Finally, in Section \ref{sec:conclusions} we present our conclusions.

\section{Effective theory}
\label{sec:eff_theory}

We consider first the effective theory describing the decay $\mu\rightarrow e\chi$, where $\chi$ is a light gauge boson, $m_\chi\ll m_\mu$. We denote the four-momenta of the muon, electron and $\chi$ as $p_\mu$, $p_e$ and $p_\chi$, respectively. The transition amplitude is given by $M=\overline{u}(p_e)\Gamma^\alpha(p_\mu,p_e) u(p_\mu) \epsilon^*_\alpha(p_\chi)$, where $\Gamma^\alpha(p_\mu,p_e)$ can be written in terms of six dimensionless scalar form factors $F_i(p_\chi^2)$, $G_i(p_\chi^2)$, $i=1,2,3$, as:
\begin{equation}
 \begin{split}
  \Gamma^\alpha=&\left(\gamma^{\alpha}-\frac{\slashed{p}_{\chi}p_\chi^\alpha}{p_\chi^2}\right) F_1(p_\chi^2)+i\frac{\sigma^{\alpha\beta}p_{\chi\beta}}{m_\mu+m_e}F_2(p_\chi^2)
+\frac{2 p_\chi^\alpha}{m_\mu+m_e} F_3(p_\chi^2)+\\
&\left(\gamma^{\alpha}-\frac{\slashed{p}_{\chi}p_\chi^\alpha}{p_\chi^2}\right)\gamma^5 G_1(p_\chi^2)+i\frac{\sigma^{\alpha\beta}\gamma^5 p_{\chi\beta}}{m_\mu+m_e}G_2(p_\chi^2)
+\frac{2 p_\chi^\alpha}{m_\mu+m_e}\gamma^5 G_3(p_\chi^2)\,,
 \end{split}
 \label{eq:Gamma_eff}
\end{equation}
where $\sigma^{\alpha\beta}=-\frac{i}{2}[\gamma^\alpha,\gamma^\beta]$.
For the decay process $\mu\rightarrow e\chi$, where the gauge boson $\chi$ is on-shell, $p_\chi^2=m_\chi^2$. The conservation of the 
$U(1)_\chi$ charge requires the form factor $F_3(p_\chi^2)$ to vanish. Moreover the Ward identities imply that $p_\chi^\alpha\cdot\epsilon^{*}_\alpha(p_\chi)=0$, so the terms proportional to $G_3(p^2_{\chi})$ and $\frac{\slashed{p}_{\chi}p_\chi^\alpha}{p_\chi^2}$ will not contribute to the amplitude. 

The decay rate can then be expressed in terms of four form factors and reads:
\begin{align}
  \Gamma(\mu\to e\chi)&=\frac{\lambda^{1/2}[m^2_\mu,m^2_e,m^2_\chi]}{16\pi m_\mu}\left[\left(1-\frac{m_e}{m_\mu}\right)^2\left(1-\frac{m_\chi^2}{(m_\mu-m_e)^2}\right)\left(2\Big|F_1(m_\chi^2)-F_2(m_\chi^2)\Big|^2\right.\right. \nonumber \\
 &\left.+\Big|F_1(m_\chi^2)\frac{(m_\mu+m_e)}{m_\chi}-F_2(m_\chi^2)\frac{m_\chi}{(m_\mu+m_e)}\Big|^2\right)+\left(1+\frac{m_e}{m_\mu}\right)^2\left(1-\frac{m_\chi^2}{(m_\mu+m_e)^2}\right)\nonumber\\ &\left.\left(2\Big|G_1(m_\chi^2)-G_2(m_\chi^2)\frac{(m_\mu-m_e)}{(m_\mu+m_e)}\Big|^2+\Big|G_1(m_\chi^2)\frac{(m_\mu-m_e)}{m_\chi}+G_2(m_\chi^2)\frac{m_\chi}{(m_\mu+m_e)}\Big|^2\right)\right]\,,
 \label{eq:rate_general}
 \end{align}
 where $\lambda[m_\mu^2,m_e^2,m_\chi^2]$ is the usual K\"all\'en function. Analogous expressions hold for the decays $\tau\rightarrow\mu\chi$ and $\tau\rightarrow e\chi$, with the appropriate substitutions. 

The term proportional to $1/m_\chi$ corresponds to the emission of the longitudinal component of the vector boson, and apparently makes the limit $m_\chi\rightarrow 0$ divergent and not continuously matched to the well studied result from  $\mu\rightarrow e\gamma$\cite{Petcov:1976ff, Marciano:1977wx, Lee:1977tib}. Therefore, in an effective field theory approach, great care should be taken when considering decays into ultralight gauge bosons, since in a gauge invariant and renormalizable theory one generically expects the rate of $\mu\rightarrow e\chi$ to be finite. 

In the following  sections we present two explicit models  where the lepton flavor violating effective interaction Eq.~(\ref{eq:Gamma_eff}) is generated either at tree-level or at the one-loop level. Apart from the interest of the models in themselves, we will show explicitly that the form factors $F_1(m_\chi^2)$ and $G_1(m_\chi^2)$ contain an implicit dependence on $m_\chi$, rooted in gauge invariance, so that the rate for $\mu\rightarrow e\chi$ is finite in the limit $m_\chi\rightarrow 0$. 

\section{$\mu\rightarrow e\chi$ at tree level}
\label{sec:tree-level}

The particle content of the model, and the corresponding spins and charges under $SU(2)_L\times U(1)_Y\times U(1)_\chi$, are summarized in Table  \ref{tab:tree_level}~\footnote{The model can be made anomaly-free adding heavy particles with suitable charges, without modifying the discussion that follows.}. Here, $L_i=(\nu_{L_i}, e_{L_i})$ and $e_{R_i}$, $i=1,2$, denote the Standard Model $SU(2)_L$ lepton doublets and singlets, respectively (we have restricted ourselves to the two generation case, although the extension to three generations is straightforward).
They have a generation independent hypercharge, $Y_L$ and $Y_e$ respectively, although we allow for generation dependent charges under $U(1)_\chi$. Further, $\phi_{jk}$, $i,j=1,2$ denote complex scalar fields, doublets under $SU(2)_L$, with hypercharge $Y_{jk}$ and charge under $U(1)_\chi$ equal to $q_{\phi_{jk}}$. 

\begin{table}[t!]
\begin{center}
\begin{tabular}{ |c| c c c c | c c c c|}
  \hline
  & $L_1$ & $L_2$ & $e_{R_1}$ & $e_{R_2}$ & $\phi_{11}$& $\phi_{12}$& $\phi_{21}$& $\phi_{22}$ \\ 
  \hline
  spin & 1/2 & 1/2 & 1/2 & 1/2 & 0 & 0 & 0 & 0\\
 $SU(2)_L$ & 2 & 2 & 1  & 1 & 2 & 2 & 2 & 2 \\
 $U(1)_Y$ & $-1/2$ & $-1/2$ & $-1$ & $-1$ & $Y_{11}$ &  $Y_{12}$ & $Y_{21}$ &  $Y_{22}$\\
 $U(1)_\chi $ & $q_{L_1}$ & $q_{L_2}$ & $q_{e_1}$ & $q_{e_2}$ & $q_{\phi_{11}}$& $q_{\phi_{12}}$& $q_{\phi_{21}}$& $q_{\phi_{22}}$\\
 \hline
\end{tabular}
\end{center}
\caption{Spins and charges under $SU(2)_L\times U(1)_Y\times U(1)_\chi$ of the particles of the model described in Section \ref{sec:tree-level}, leading to the decay $\mu\rightarrow e\chi$ at tree level. All fields are assumed to be singlets under $SU(3)_C$.}
\label{tab:tree_level}
\end{table}

The kinetic terms of the particles of the model read:
\begin{align}
{\cal L}_{\rm kin}=\sum_{j=1}^2 i( \overline L_j\slashed{D} L_j+ \overline e_{R_j} \slashed{D} e_{R_j})+\sum_{j,k=1}^2 (D_\mu \phi_{jk})^\dagger (D^\mu \phi_{jk} )\;,
\end{align}
where $D_\mu$ denotes the covariant derivative, given by
\begin{align}
D_\mu&=\partial_\mu +i g W_\mu^a T_a+i g' Y B_\mu + i g_\chi q \chi_\mu~~{\rm for~the }~SU(2)_L~{\rm doublets}\;, \nonumber \\
D_\mu&=\partial_\mu +i g' Y B_\mu + i g_\chi q \chi_\mu ~~{\rm for~the}~SU(2)_L~{\rm singlets}\;,
\label{eq:L_kin_tree_level}
\end{align}
with $g$, $g'$ and $g_\chi$ the coupling constants of $SU(2)_L$, $U(1)_Y$ and $U(1)_\chi$ respectively. 

We also assume $Y_{jk}=1/2$. Then, for $j,k$ such that $q_{\phi_{jk}}=q_{L_j}-q_{e_k}$ the following Yukawa couplings arise in the Lagrangian:
\begin{align}
-{\cal L}_{\rm Yuk}=\sum_{j,k=1}^2 y_{jk} \overline L_j \phi_{jk} e_{R_k}+{\rm h.c.}
\end{align}
We also assume that the doublet scalars acquire a vacuum expectation value for some $i,j$, $\langle \phi_{jk}\rangle =v_{jk}$. To keep the discussion general, we consider that the charges of the particles allow all Yukawa couplings, and that all $\phi_{jk}$ acquire a vacuum expectation value; the different subcases follow straightforwardly by setting the corresponding $y_{jk}$ and/or $v_{jk}$ to zero. 

The non-zero expectation values for $\phi_{jk}$ generate a mass for the $\chi$ boson: 
\begin{align}
m^2_\chi=g_\chi^2 (q_{\phi_{11}}^2 v_{11}^2+q_{\phi_{12}}^2 v_{12}^2+q_{\phi_{21}}^2 v_{21}^2+q_{\phi_{22}}^2 v_{22}^2)\;.
\label{eq:chi-mass}
\end{align}
Furthermore, since $\phi_{jk}$ have charge under $SU(2)_L\times U(1)_Y$, their expectation value would also contribute to the $Z$ and $W$ masses. Since we are assuming $m_\chi\ll m_\mu$, this contribution can be safely neglected.

The expectation value of the doublet scalars generates a mass term for the charged leptons, $-{\cal L}_{\rm mass}\supset \overline{e_{L_j}} M_{jk} e_{R_k}+{\rm h.c.}$, with 
\begin{align}
M=\begin{pmatrix}
y_{11} v_{11} & y_{12} v_{12} \\
y_{21} v_{21} & y_{22} v_{22} \\
 \end{pmatrix}\;.
 \label{eq:mass_matrix}
\end{align}
We now rotate the fields to express the Lagrangian in the mass eigenstate basis:
\begin{align}
\begin{pmatrix} e_L \\ \mu_L \end{pmatrix}= \begin{pmatrix} \cos\theta_L & \sin\theta_L \\-\sin\theta_L & \cos\theta_L \end{pmatrix} \begin{pmatrix} e_{L_1} \\ e_{L_2} \end{pmatrix}\,,~~~~~~~
\begin{pmatrix} e_R \\ \mu_R \end{pmatrix}= \begin{pmatrix} \cos\theta_R & \sin\theta_R \\-\sin\theta_R & \cos\theta_R \end{pmatrix} \begin{pmatrix} e_{R_1} \\ e_{R_2} \end{pmatrix}
\end{align}
so that $-{\cal L}_{mass}\supset \overline{e_{L}} m_e e_R +\overline{\mu_{L}} m_\mu \mu_R\,+$ h.c., with 
\begin{align}
m_\mu^2&\simeq y_{11}^2v_{11}^2+y_{12}^2v_{12}^2 +y_{21}^2v_{21}^2 +y_{22}^2v_{22}^2 \;,\nonumber \\
m_e^2&\simeq \frac{(y_{11} v_{11} y_{22} v_{22}-y_{12} v_{12} y_{21} v_{21})^2}{y_{11}^2v_{11}^2+y_{12}^2v_{12}^2 +y_{21}^2v_{21}^2 +y_{22}^2v_{22}^2}\;, \nonumber\\
\sin2\theta_L&\simeq-2\frac{y_{11} v_{11} y_{21} v_{21}+y_{12} v_{12} y_{22} v_{22}}{y_{11}^2v_{11}^2+y_{12}^2v_{12}^2 +y_{21}^2v_{21}^2 +y_{22}^2v_{22}^2}\;,\nonumber\\
\sin2\theta_R&\simeq -2\frac{y_{11} v_{11} y_{12} v_{12}+y_{21} v_{21} y_{22} v_{22}}{y_{11}^2v_{11}^2+y_{12}^2v_{12}^2 +y_{21}^2v_{21}^2 +y_{22}^2v_{22}^2}\;,
\label{eq:eigensystem}
\end{align}
where we have used that empirically $m_\mu\gg m_e$.

Finally, we recast the kinetic Lagrangian Eq.~(\ref{eq:L_kin_tree_level}) in terms of the mass eigenstates. We find flavor violating terms of the form 
\begin{align}
-{\cal L}\supset \overline{e_R} i g_{e \mu}^{RR} \gamma^\rho \chi_\rho  \mu_R +\overline{e_L} i g_{e \mu}^{LL} \gamma^\rho \chi_\rho \mu_L +{\rm h.c.}\;,
\label{eq:LFV-tree-level-model}
\end{align}
with
\begin{align}
g_{e\mu}^{RR}&=g_\chi (q_{e_{1}}-q_{e_{2}})\sin\theta_R\cos\theta_R \;,\nonumber \\
g_{e\mu}^{LL}&=g_\chi (q_{L_{1}}-q_{L_{2}})\sin\theta_L\cos\theta_L\;.
\end{align}
Clearly, if the $U(1)_\chi$ charges are generation independent, the flavor violation is absent at tree-level (as is the case for the photon and $Z$  flavor violating couplings). Further, if the interaction eigenstates are aligned to the mass eigenstates, the tree-level flavor violation is also absent. 

Comparing to the general form of the lepton flavor violating interaction vertex, Eq.~(\ref{eq:Gamma_eff}), one can identify
\begin{align}
F_1=\frac{1}{2}(g_{e\mu}^{RR}+g_{e\mu}^{LL}) \;,\nonumber\\
G_1=\frac{1}{2}(g_{e\mu}^{RR}-g_{e\mu}^{LL}) \;,
\end{align}
while all other form factors vanish at tree-level. The rate for $\mu\rightarrow e\chi$  then reads:
\begin{align}
  \Gamma(\mu\to e\chi)&=\frac{m_\mu}{32\pi}\Big(\big|g_{e\mu}^{LL}\big|^2+\big|g_{e\mu}^{RR}\big|^2\Big)\left(2+\frac{m_\mu^2}{m_\chi^2}\right)\left(1-\frac{m_\chi^2}{m_\mu^2}\right)^{2}\;,
  \label{eq:mutoechi_TLM}
 \end{align}
where we have neglected the electron mass against the muon mass. Naively, the term $m_\mu^2/m_\chi^2$ would enhance the rate as $m_\chi\rightarrow 0$. However, if the gauge and fermion masses arise as a consequence of the spontaneous breaking of the $U(1)_\chi$ symmetry,  the limit $m_\chi\rightarrow 0$ requires $v_{jk}\rightarrow 0$ for all $i,j$, which in turn implies $m_\mu\rightarrow 0$. One can explicitly check from Eqs.~(\ref{eq:chi-mass}) and (\ref{eq:eigensystem}) that indeed when $m_\chi\rightarrow 0$ the term $m_\mu^2/m_\chi^2$ is finite
(as expected from the Goldstone boson equivalence theorem \cite{Cornwall:1974km, Vayonakis:1976vz, Lee:1977eg, Chanowitz:1984ne}), and depends on a function of the Yukawa couplings, the gauge coupling, and the charges and vacuum expectation values of the fields $\phi_{jk}$.~\footnote{An analogous behaviour occurs in the  top decay $t\rightarrow b W^+$. The decay rate is  $\Gamma(t\rightarrow b W^+)\sim m_t^3/m_W^2$ and naively diverges when $m_W\rightarrow 0$. However, since both masses arise as a consequence of the spontaneous breaking of the electroweak symmetry, $\Gamma(t\rightarrow b W^+)\sim m_t y_t^2/g^2$ and is finite.}

For example, for the specific case where the Yukawa couplings satisfy $y_{22}  \gg y_{11}\gg  y_{12} ,y_{21}$ and all $v_{jk}= v$, so that the mass matrix $M$ is almost diagonal, and $q_{jk}=Q$, the relevant parameters of the model after the spontaneous  breaking of the gauge symmetries are:
\begin{equation}
\begin{gathered}
m_\mu^2\simeq y_{22}^2v^2, ~~~~~
m_e^2 \simeq y_{11}^2 v^2, ~~~~~
m_\chi^2 \simeq 4g_\chi^2 Q^2 v^2\;, \\
\sin2\theta_L\simeq-2\frac{y_{12}  }{ y_{22} } , ~~~~~
\sin2\theta_R \simeq -2\frac{y_{21}  }{y_{22}}\;.
\end{gathered}
\label{eq:example_TLM}
\end{equation}
Therefore, the rate for $\mu\to e\chi$ in the limit $m_\chi\rightarrow 0$ is given by
\begin{align}
  \Gamma(\mu\to e\chi)\Big|_{m_\chi\rightarrow 0}&\simeq \frac{m_\mu}{32\pi}\frac{g_\chi^2}{y^2_{22}}\left(2+\frac{y_{22}^2}{4 g_\chi^2 Q^2}\right)\left(1-\frac{4g_\chi^2 Q^2}{y_{22}^2}\right)^{2}\Big[y^2_{12}(q_{L_{1}}-q_{L_{2}})^2+y^2_{21}(q_{e_{1}}-q_{e_{2}})^2  \Big]\;.
 \end{align}
The rate is maximal when $Q g_\chi/y_{22}\to 0$, and zero when  $Q g_\chi/y_{22}\simeq 1/2$, which corresponds to  $m_\chi/m_\mu\simeq 1$, {\it i.e.} when the phase space available for the decay closes.  For most values of $g_\chi/y_{22}$, the prefactor is $\sim 10^{-3}$, and therefore the rate can only be suppressed by invoking small couplings $y_{12}, y_{21}$, or by postulating intergenerational universality of the $U(1)_\chi$ charges. In the latter case, the process $\mu\rightarrow e\chi$ could be generated at the one loop level, as we will discuss in the next section. 

A complementary probe of the $\mu$-$e$ flavor violation is the three-body decay $\mu^+\rightarrow e^+ e^- e^+$, which is generated in this model at tree-level via the exchange of a virtual $\chi$. The flavor conserving interaction vertex of the electron with the $\chi$-boson has a similar form as Eq. (\ref{eq:LFV-tree-level-model}):
\begin{align}
-{\cal L}\supset \overline{e_R} i g_{e e}^{RR} \gamma^\rho \chi_\rho  e_R +\overline{e_L} i g_{e e}^{LL} \gamma^\rho \chi_\rho  e_L \;,
\label{eq:LFC-tree-level-model}
\end{align}
where 
\begin{align}
g_{ee}^{RR}&=g_\chi \left(q_{e_{2}}\sin^2\theta_R+q_{e_{1}}\cos^2\theta_R \right)\;,\nonumber \\
g_{ee}^{LL}&=g_\chi \left(q_{L_{2}}\sin^2\theta_L+q_{L_{1}}\cos^2\theta_L\right)\;.
\end{align}
The doubly differential decay width for $\mu\to 3e$ can be calculated from the interaction Lagrangians Eqs.~(\ref{eq:LFC-tree-level-model}) and~(\ref{eq:LFV-tree-level-model}) and reads:
\begin{align}
 \frac{\dd^2 \Gamma(\mu\to3e)}{\dd s~\dd t}&=\frac{1}{128 \pi^3 m_\mu^3}\Bigg[ (|g_{ee}^{LL}|^2 |g_{e\mu}^{RR}|^2+|g_{ee}^{RR}|^2 |g_{e\mu}^{LL}|^2) \left(\frac{t(m_\mu^2-t)}{(m_\chi^2-s)^2+m_\chi^2\Gamma_\chi^2}+\frac{s(m_\mu^2-s)}{(m_\chi^2-t)^2+m_\chi^2\Gamma_\chi^2}\right)\nonumber\\ &+\Big(|g_{ee}^{LL}|^2 |g_{e\mu}^{LL}|^2+|g_{ee}^{RR}|^2 |g_{e\mu}^{RR}|^2\Big)\left(\frac{(s+t) \left(m_\mu^2-s-t\right)\big(4\Gamma^2_\chi m^2_\chi+(s+t-2m_\chi^2)^2\big)}{((m_\chi^2-s)^2+m_\chi^2\Gamma_\chi^2)((m_\chi^2-t)^2+m_\chi^2\Gamma_\chi^2)}\right)
 \Bigg]\;,
 \label{eq:doubly_differential_TLM}
\end{align}
where we have defined $s\equiv (p_\mu-p_{e_1})^2$ and $t\equiv (p_\mu-p_{e_2})^2$, with $p_{e_1}$ and $p_{e_2}$ the electron momenta, and which are kinematically restricted to be in the range:
\begin{equation}
 0\leq t \leq (m_\mu^2-s) \quad \text{and} \quad 0\leq s\ \leq m^2_\mu\,.
 \label{eq_integration_limits}
\end{equation}
Further, $\Gamma_\chi$ is the total width of the $\chi$-boson.
 We focus in what follows in a scenario where $1~\text{MeV} \lesssim m_\chi \lesssim m_\mu$. In this mass range, the dominant decay channels are  $\chi\rightarrow e^-e^+, \overline{\nu_{L_1}}\nu_{L_1}, \overline{\nu_{L_2}}\nu_{L_2}$.
Using the electron interaction vertex from Eq.~(\ref{eq:LFC-tree-level-model}) and the neutrino interaction vertex from Eq.~(\ref{eq:L_kin_tree_level}), we find that the total decay width is:
\begin{equation}
\Gamma_\chi=\frac{m_\chi}{24\pi}\left(|g_{ee}^{LL}|^2+|g_{ee}^{RR}|^2+|g_{\chi}q_{L_1}|^2+|g_{\chi}q_{L_2}|^2\right)\,,
\label{eq:chi-width}
\end{equation}
which satisfies $\Gamma_\chi\ll m_\chi$. 

\begin{figure}[t!]
\centering
\includegraphics[width=0.5\linewidth]{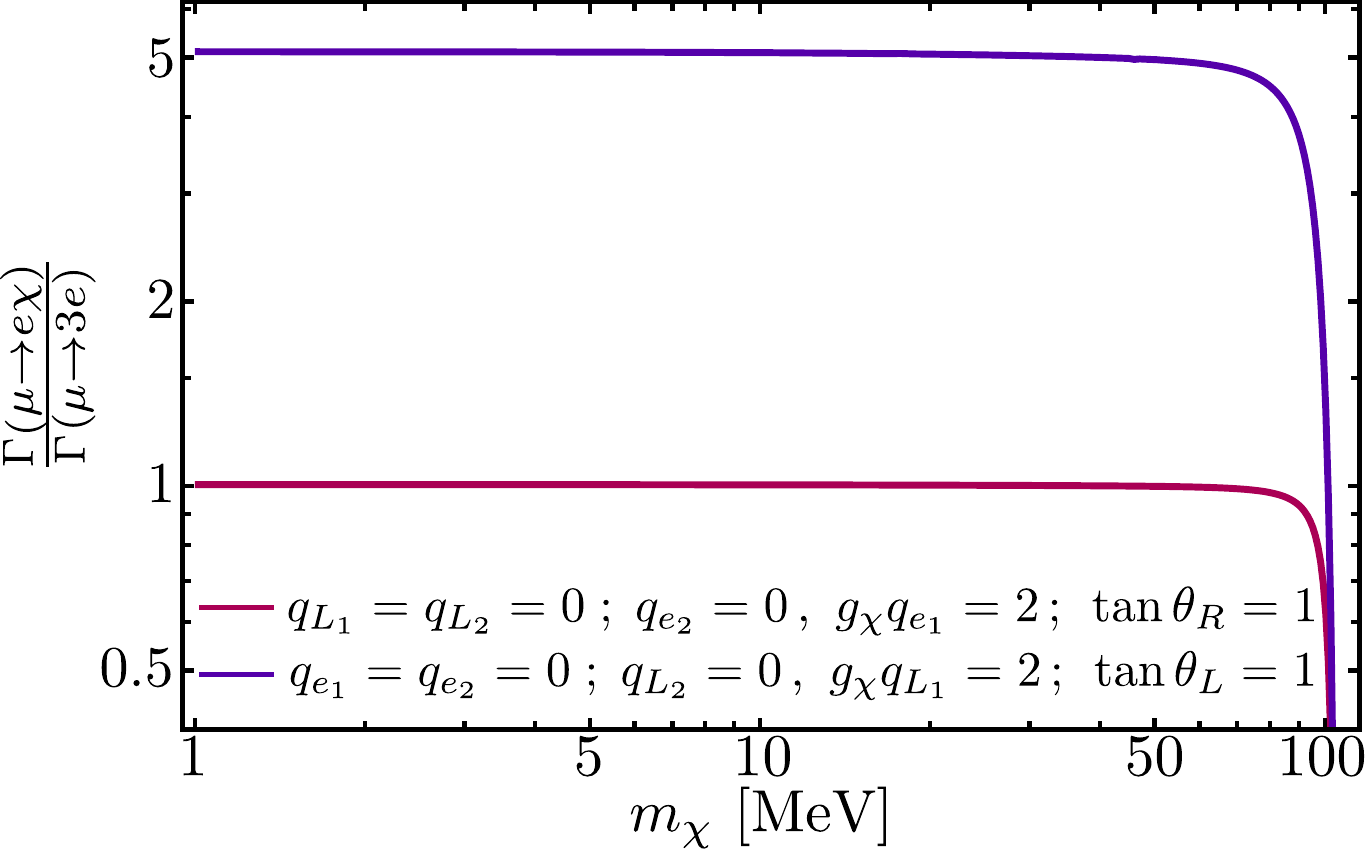}
\caption{Ratio of rates of $\mu\to e\chi$ and $\mu\to 3e$ as a function of $m_\chi$ for the tree-level model presented in Section \ref{sec:tree-level}, for the cases described in the text $q_{L_i}=0$, $q_{e_2}=0$, $g_\chi q_{e_1}=2$, and $\tan\theta_R=1$ (magenta line); and $q_{e_i}=0$, $q_{L_2}=0$, $g_\chi q_{L_1}=2$, and $\tan\theta_L=1$ (blue line).}
\label{fig:ratio-TLM}
\end{figure}

We show in Fig.~\ref{fig:ratio-TLM} the ratio between $\Gamma(\mu\to e\chi)$ and $\Gamma(\mu\to 3e)$ as a function of  $m_\chi$,  for a representative case where $\chi$ couples only to the right-handed leptons ({\it i.e.} $q_{L_1}=q_{L_2}=0$; and $q_{e_2}=0$, $g_\chi q_{e_1}=2$, and $\tan\theta_R=1$)
or when $\chi$ couples only to the  left-handed leptons ({\it i.e.} $q_{e_1}=q_{e_2}=0$; and $q_{L_2}=0$, $g_\chi q_{L_1}=2$, and $\tan\theta_L=1$).
In the former case,  the ratio is $\simeq 1$, and in the latter it is $\simeq 5$; with a mild sensitivity to the concrete choices of the charges and mixing angles. This result can be understood using the narrow width approximation (NWA), which holds when $\chi$ is produced close to the mass shell. Under this approximation, one can replace in the propagators:
\begin{equation}
 \frac{1}{(x-m_\chi^2)^2+m_\chi^2 \Gamma_\chi^2}\to \frac{\pi}{m_\chi \Gamma_\chi}\delta(x-m_\chi^2)\,,
\end{equation}
where $x$ is any Mandelstam variable. 
Under this approximation, the decay rate for $\mu\to3e$ reads:
\begin{align}
 \Gamma(\mu\to3e)&\simeq \frac{m_\mu}{32\pi}\frac{\left(|g_{e\mu}^{LL}|^2+|g_{e\mu}^{RR}|^2\right)\left(|g_{ee}^{LL}|^2+|g_{ee}^{RR}|^2\right)}{|g_{ee}^{LL}|^2+|g_{ee}^{RR}|^2+|g_{\chi}q_{L_1}|^2+|g_{\chi}q_{L_2}|^2}\left(2+\frac{m_\mu^2}{m_\chi^2}\right)\left(1-\frac{m_\chi^2}{m_\mu^2}\right)^2\nonumber\\ &+\frac{m_\chi}{64\pi}\Big(|g_{ee}^{LL}|^2|g_{e\mu}^{LL}|^2+|g_{ee}^{RR}|^2|g_{e\mu}^{RR}|^2\Big)\frac{m_\chi}{m_\mu}\left(1-2\frac{m_\chi^2}{m_\mu^2}\right)\,.
 \label{eq:Rate-3e-TLM-NWA}
\end{align}

As for the decay $\mu\rightarrow e\chi$ the rate apparently diverges as $m_\chi\rightarrow 0$, but is in fact finite since  $m_\mu$ and $m_\chi$ are both generated after the breaking of the $U(1)_\chi$ symmetry. Further, and using Eq.~(\ref{eq:mutoechi_TLM}), one reproduces the result $\Gamma(\mu\rightarrow e\chi)/\Gamma(\mu\rightarrow 3e)\simeq 1 $ or $\simeq 5$ that we obtained numerically for our two representative scenarios. 
As $m_\chi$ becomes larger, the ratio becomes sensitive to the underlying model parameters, although this sensitivity is suppressed by a factor $m_\chi^2/m_\mu^2$, and is hence typically weak, in agreement with the numerical results of Fig.~\ref{fig:ratio-TLM}.

Given the current experimental limits on ${\rm Br}(\mu\rightarrow e \chi)$ and ${\rm Br}(\mu\rightarrow 3 e)$, the most stringent constraints on the model will stem from the latter process~\footnote{We note that the on-shell $\chi$ decays into $e^+e^-$  with a decay length $L_\chi\simeq7.9\times10^{-10}\mathrm{m}\,(m_\chi/ \mathrm{MeV})^{-2}g^{-2}$,  where $g$ is a combination of couplings, {\it cf.} Eq.~(\ref{eq:chi-width}), and therefore the decay occurs inside the detector.}. For the case $m_\chi\ll m_\mu$, and using the upper limit  ${\rm Br}(\mu\to 3e)\leq 1.0\times 10^{-12}$ from SINDRUM, one finds very stringent constraints on the strength of the effective couplings. Concretely, when  $q_{L_i}=0$, one finds $|g_{e\mu}^{RR}|/m_\chi \lesssim 1.6\times 10^{-16}/{\rm MeV}$. 
This limit on the effective parameters can in turn be translated into limits on the Yukawa couplings, $U(1)_\chi$-charges and gauge coupling, and expectation values of the scalar doublets $\phi_{ij}$, with the restriction of reproducing the correct muon mass $m_\mu\simeq 105$ MeV.

\section{$\mu\rightarrow e\chi$ at the one loop level}
\label{sec:one-loop}

In this section we present a renormalizable model with generation independent $U(1)_\chi$ charges, and containing new fields that generate the process $\mu\to e\chi$ at the one loop level.
To provide masses to charged leptons we introduce a doublet scalar, with hypercharge $+1/2$, and with $U(1)_\chi$-charge $q_\phi=q_L-q_e$, such that the Yukawa coupling $y_{jk} \overline L_j  e_{R_k} \phi+\mathrm{h.c.}$ is allowed. This choice leads to the conservation of the electron and the muon flavors, akin to the Standard Model. To violate the lepton flavor numbers, we introduce a new Dirac fermion $\psi$ and a new complex scalar $\eta$, both singlets under $SU(2)_L$, with hypercharges and $Y_\psi$ and $Y_\eta$, and $U(1)_\chi$-charges $q_\psi$ and $q_\eta$ respectively. The spins and charges of the particles of the model are listed in Table \ref{tab:one-loop}.

\begin{table}[t!]
\begin{center}
\begin{tabular}{ |c| c c c c | c c c |}
  \hline
  & $L_1$ & $L_2$ & $e_{R_1}$ & $e_{R_2}$ & $\phi$  & $\psi$ & $\eta$\\
  \hline
spin & 1/2 & 1/2 & 1/2 & 1/2 & 0 & 1/2 & 0 \\
 $SU(2)_L$ & 2 & 2 & 1 & 1 & 2 & 1 & 1 \\
 $U(1)_Y$ & $-1/2$ & $-1/2$ & $-1$ & $-1$ & $+1/2$ & $Y_\psi$ & $Y_\eta$\\
 $U(1)_\chi $ & $q_{L}$ & $q_{L}$ & $q_{e}$ & $q_{e}$& $q_{\phi}$ &$q_\psi$ & $q_\eta$\\
  \hline
\end{tabular}
\end{center}
\caption{Spins and charges under $SU(2)_L\times U(1)_Y\times U(1)_\chi$ of the particles of the model described in Section \ref{sec:one-loop}, leading to the decay $\mu\rightarrow e\chi$ at the one loop level. All fields are assumed to be singlets under $SU(3)_C$.}
\label{tab:one-loop}
\end{table}

We assume that $q_e=q_\psi+q_\eta$ and $Y_e=Y_\psi+Y_\eta$, so that the Yukawa couplings $y^\prime_i \overline e_{R_i} \psi\eta$ are allowed by the symmetries. We also assume that $\phi$ acquires a vacuum expectation value, but $\eta$ does not. The gauge boson of the $U(1)_\chi$ symmetry then acquires a mass
\begin{align}
m_\chi=g_\chi q_\phi \langle \phi\rangle\;, 
\end{align}
which we assume to be $m_\chi\ll m_\mu$.~\footnote{In this simple model, $m_\mu$, $m_e$ and $m_\chi$ are all proportional to $\langle \phi\rangle$. However, one can completely uncorrelate the fermion masses and the gauge boson masses by imposing $q_\phi=0$ and by postulating the existence of another scalar field, whose expectation value contributes to $m_\chi$, but not to the fermion masses.} Further, a mass matrix for the charged leptons is generated, of the form Eq.~(\ref{eq:mass_matrix}). Let us note that if $\eta$ acquires an expectation value, a mixing between $e_{R_i}$ and $\psi$ is generated, and the mass matrix becomes instead $3\times 3$. The analysis in that case would be analogous, although we disregard that possibility for simplicity and assume that $\langle \eta\rangle=0$.

Recasting the Lagrangian in terms of the mass eigenstates of the theory, $e_{L,R}$ and $\mu_{L,R}$, one finds interaction terms  with the massive gauge boson $\chi$ of the form 
\begin{eqnarray}
 \mathcal{L}\supset& & -ig_\chi q_L\left(\overline{e_L}\gamma^\nu e_L+\overline{\mu_L}\gamma^\nu \mu_L+\overline{\nu_{L1}}\gamma^\nu \nu_{L1}+\overline{\nu_{L2}}\gamma^\nu \nu_{L2}\right)\chi_\nu -i g_\chi q_e\left(\overline{e_R}\gamma^\nu e_R+\overline{\mu_R}\gamma^\nu \mu_R\right)\chi_\nu  \nonumber\\
 && -i g_\chi q_\Psi  \overline{\Psi} \gamma^\nu \Psi \chi_\nu-iq_\eta g_\chi \left[\eta^*(\partial^\nu \eta)-(\partial^\nu\eta^*)\eta\right]\chi_\nu
\,,
\end{eqnarray}
as well as a Yukawa coupling to the right-handed leptons:
\begin{equation}
 \mathcal{L}\supset h_e \overline{e_R} \eta\psi+ h_\mu \overline{\mu_R} \eta\psi +{\rm h.c.}
\end{equation}

\begin{figure}[t!]
	\centering
	\includegraphics[width=168mm]{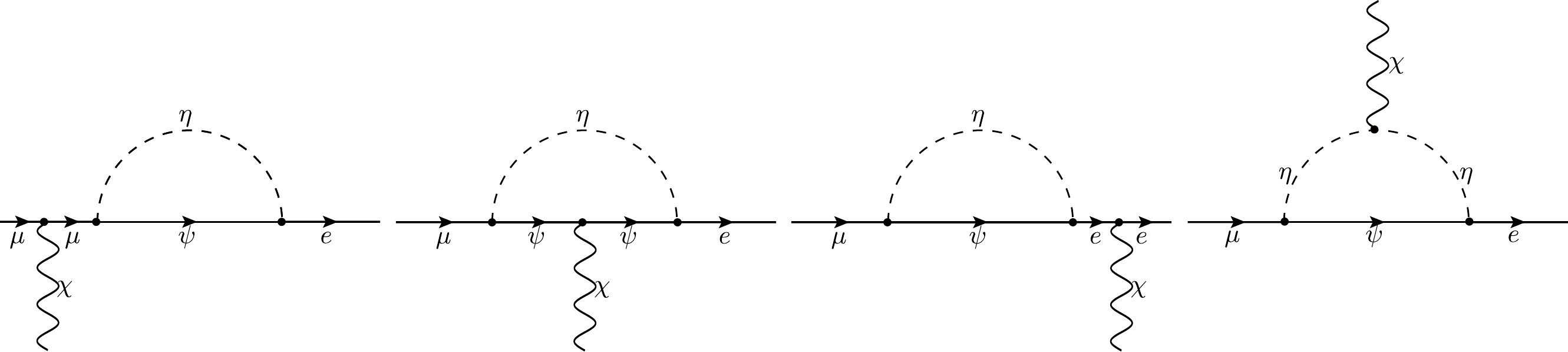} 
	\caption{One loop diagrams contributing to the decay $\mu\to e\chi$.}
	\label{Diagrams}
\end{figure}

\begin{figure}[t!]
	\centering
	\includegraphics[width=82mm]{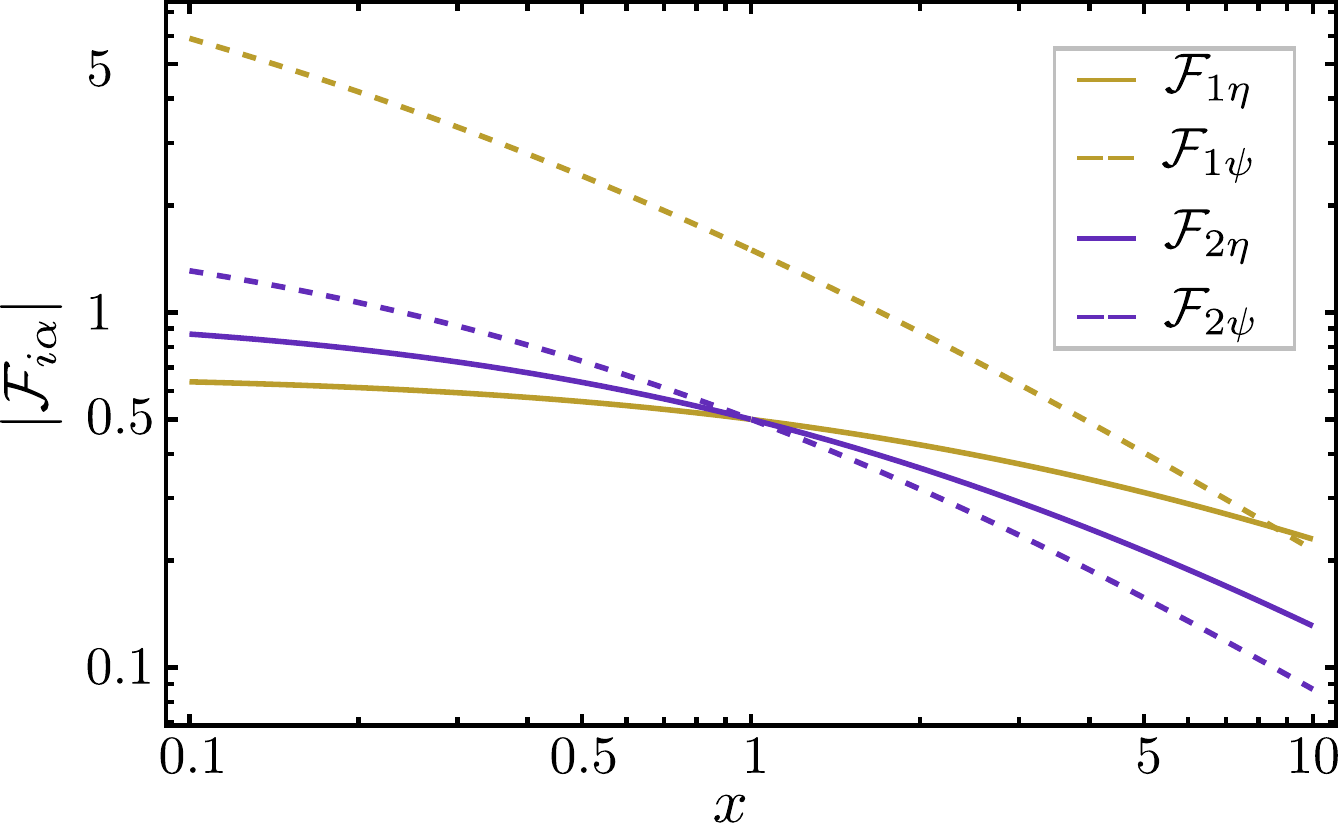} 
	\caption{Moduli of the functions ${\cal F}_{i\alpha}(x)$, with $i=1,2$ and $\alpha=\eta,\psi$, as defined in Eq.~(\ref{F2-psi}).}
	\label{FF-}
\end{figure}

The process $\mu\rightarrow e\chi$ is generated in this model at the one loop-level, through the four diagrams shown in Fig. \ref{Diagrams}. The form factors  are finite and read:
\begin{align}
F_1(m_\chi^2) &=~~ G_1(m_\chi^2) = \frac{g_\chi h_e h_\mu }{384 \pi^2}\frac{m_\chi^2} {M_\eta^2} \Bigg[q_\eta {\cal F}_{1\eta}  \Big(\frac{M^2_\psi}{M^2_\eta}\Big)+q_\psi {\cal F}_{1\psi} \Big(\frac{M^2_\psi}{M^2_\eta}\Big)\Bigg]\,,   \nonumber\\
F_2(m_\chi^2) &= -G_2(m_\chi^2)=\frac{g_\chi h_e h_\mu}{384 \pi ^2}\frac{m^2_\mu}{M^2_\eta} \Bigg[q_\eta {\cal F}_{2\eta} \Big(\frac{M^2_\psi}{M^2_\eta}\Big)+q_\psi {\cal F}_{2\psi} \Big(\frac{M^2_\psi}{M^2_\eta}\Big)\Bigg]\,,\label{FF-2}
\end{align}
where 
\begin{align}
{\cal F}_{1\eta}(x)&=\frac{-2+9x-18x^2+x^3\left(11-6\ln x\right)}{3\left(1-x\right)^4}\,, \nonumber \\
{\cal F}_{1\psi}(x)&=\frac{16-45x+36x^2-7x^3+6\left(2-3x\right)\ln x}{3\left(1-x\right)^4}\,,  \nonumber\\
{\cal F}_{2\eta}(x)&=\frac{1-6x+3x^2(1-2\ln x)+2x^3}{\left(1-x\right)^4}\,, \nonumber\\
{\cal F}_{2\psi}(x)&=\frac{-2-3x(1+2\ln x)+6x^2-x^3}{\left(1-x\right)^4}\,.\label{F2-psi}
\end{align}
The absolute values of these functions are shown in Fig.~\ref{FF-}, and are regular at $x=1$.

Using Eq.~(\ref{eq:rate_general}),  and that $F_1=G_1$, $F_2=-G_2$, the decay rate can be recast as
 \begin{align}
  \Gamma(\mu\to e\chi)\simeq\frac{m_\mu}{8\pi}\left(1-\frac{m_\chi^2}{m_\mu^2}\right)^2\left[\Big|F_1(m_\chi^2)\frac{m_\mu}{m_\chi}-F_2(m_\chi^2)\frac{m_\chi}{m_\mu}\Big|^2+2\Big|F_1(m_\chi^2)-F_2(m_\chi^2)\Big|^2\right]\;,
  \label{eq:rate_loop_model}
 \end{align}
 where we have neglected the electron mass. The form factor $F_1$ (and $G_1$) is proportional to $m_\chi^2/M_\eta^2$, while $F_2$ (and $G_2$) is proportional to $m^2_\mu/M^2_\eta$. Inserting these form factors in the rate Eq.~(\ref{eq:rate_loop_model}), one finds that the factors $1/m_\chi$ from the emission of the longitudinal polarization cancel with the factors $m_\chi^2$ implicit in the form factors $F_1$ and $G_1$, yielding a finite rate for $\mu\rightarrow e\chi$ in the limit $m_\chi\rightarrow 0$. Further, one finds that $F_1 m_\mu/m_\chi \propto m_\chi m_\mu/M_\eta^2$, and $F_2 m_\chi/m_\mu\propto  m_\chi m_\mu/M^2_\eta$. Since we are assuming $M_\eta, M_\psi\gg m_\mu$, it follows that the rate in the limit $m_\chi\rightarrow 0 $ will depend mostly on the form factors $F_2$ and $G_2$, and can be well approximated by:
\begin{equation}\label{gammaT1}
 \Gamma(\mu\to e\chi)\Big|_{m_\chi\to 0}\simeq \frac{g_\chi^2~|h_e|^2 |h_\mu|^2 }{(768)^2~ \pi^5 }\frac{m_\mu^5 }{M_\eta^4} \Bigg[q_\eta{\cal F}_{2\eta}\left(\frac{M^2_\psi}{M^2_\eta}\right)+q_\psi{\cal F}_{2\psi}\left(\frac{M^2_\psi}{M^2_\eta}\right) \Bigg]^2\,.
\end{equation}
Nevertheless, the form factors $F_1$ and $G_1$ generate a sizable contribution to the rate when $m_\chi/m_\mu\gtrsim 0.1$.

\begin{figure}[t!]
\begin{center}
\includegraphics[width=0.49\linewidth]{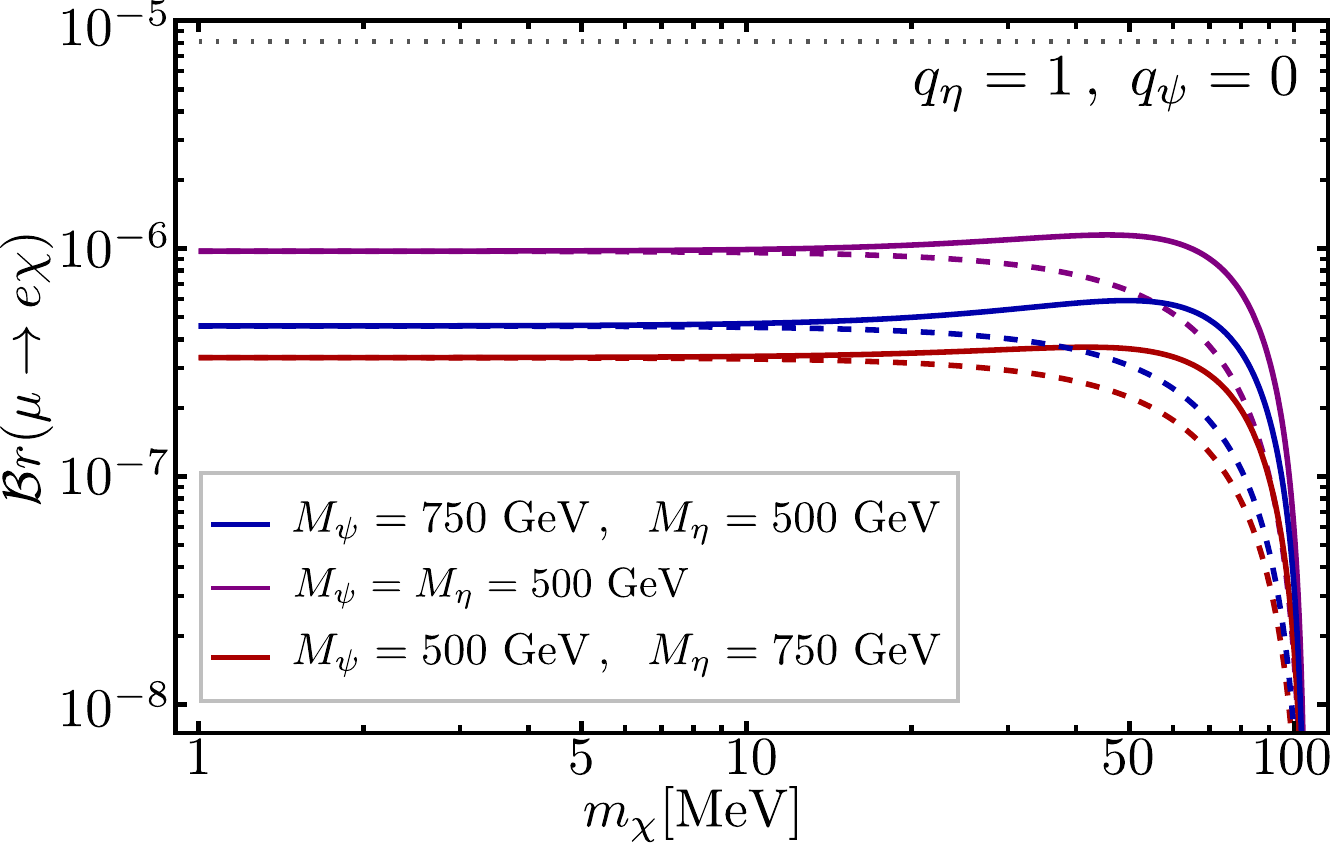}~~
\includegraphics[width=0.49\linewidth]{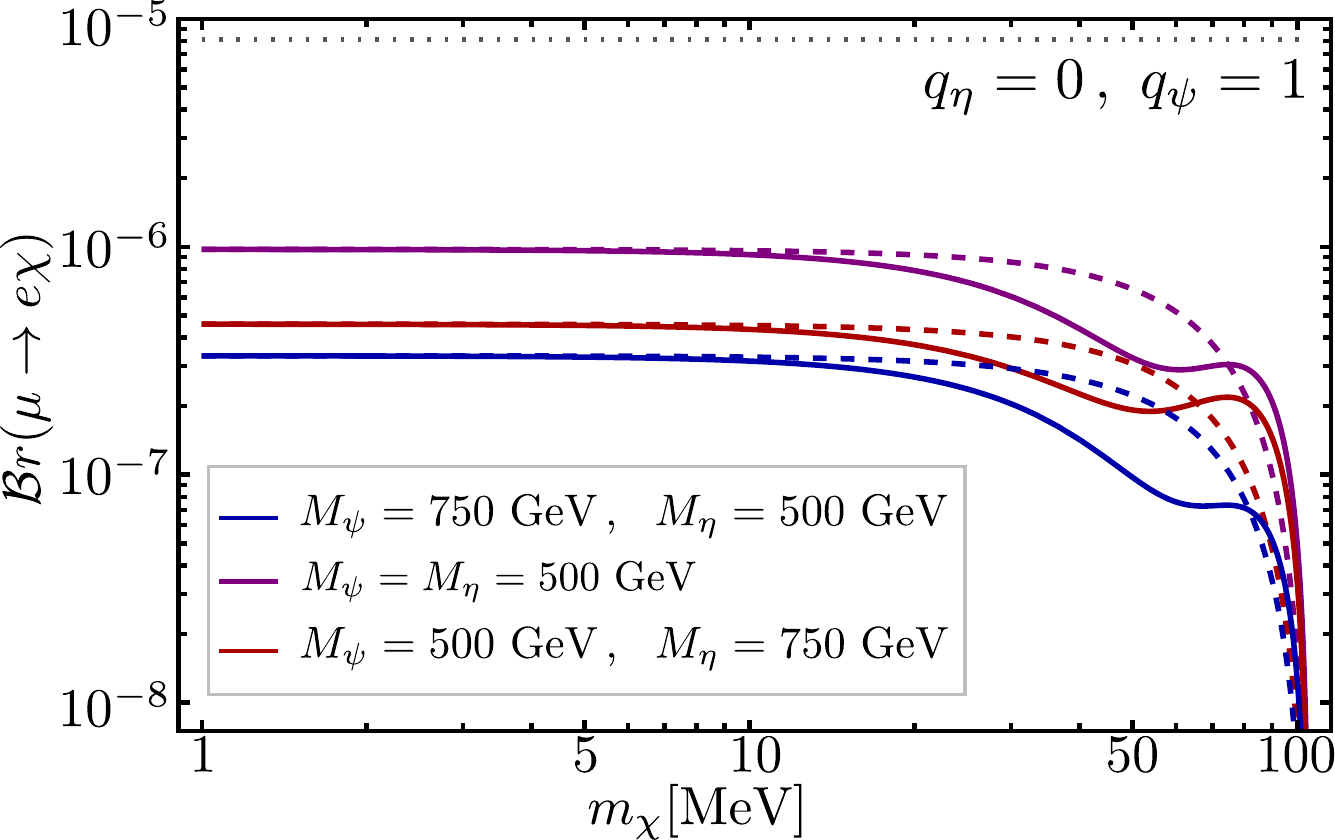}
\end{center}
\caption{Branching ratio of the process $\mu\rightarrow e\chi$ as a function of $m_\chi$ for the one loop model presented in Section \ref{sec:one-loop}, assuming $q_\eta=1$ and $q_\psi=0$ (left plot), and $q_\eta=0$ and $q_\psi=1$ (right plot); in both cases it was assumed $h_e h_\mu=1$ and $g_\chi=1$. The solid lines show the full result obtained from Eq.~(\ref{eq:rate_loop_model}), while the dashed lines neglect the contribution from $F_1$. The grey dotted line indicates the current upper limit on $Br(\mu\rightarrow e\chi)$ from the TWIST collaboration. }
\label{fig:mu_to_echi_loop}
\end{figure}

We show in Fig \ref{fig:mu_to_echi_loop},  the branching ratio for $\mu\rightarrow e\chi$ for two representative choices of charges, $q_\eta=1$ and $q_\psi=0$ (left plot) and $q_\eta=0$ and $q_\psi=1$ (right panel), and three choices of the masses of the particles in the loop: $M_\psi=750$ GeV and $M_\eta=500$ GeV (blue line), $M_\psi=M_\eta=500$ GeV (purple line), and $M_\psi=500$ GeV and $M_\eta=750$ GeV (red line). These values are compatible with the current searches for exotic charged particles \cite{Sirunyan:2018nwe,Aad:2019vnb}. We have also taken for concreteness  $h_e h_\mu=1$ and $g_\chi=1$, although the scaling of the rates with the Yukawa couplings is straightforward. The solid lines show the full result calculated using Eq.~(\ref{eq:rate_loop_model}), while the dashed lines assume $F_1=G_1=0$. As apparent for the plot, while for $m_\chi\ll m_\mu$ the form factors $F_1$ and $G_1$ can be neglected, they modify the rate when $m_\chi/m_\mu\gtrsim 0.1$, especially close to the threshold. We also show the current 90$\%$ C.L. upper limit upper limit $Br(\mu\to e\chi)< 8.1 \times 10^{-6}$ from the TWIST collaboration \cite{Bayes:2014lxz}.

The process $\mu^+ \to e^+ e^- e^+$ is generated in this toy model also at the one loop-level, through $\chi$-penguin and through box diagrams; the former are proportional to $h_e^2 h_\mu^2 g_\chi^4$ and the latter to  $h_e^6 h_\mu^2$. Assuming $h_e\ll g_\chi$, the decay will be dominated by the penguin diagrams, with doubly differential rate given by:
\begin{align}
\frac{\dd^2\Gamma(\mu\to 3e)}{\dd s~\dd t}\simeq& \frac{g_\chi^2}{32\pi^3 m_\mu^5}\Bigg[\frac{1}{\left(m_\chi^2-s\right)^2+m_\chi^2\Gamma_\chi^2}\bigg(|q_e|^2(m_\mu^2-s-t)\Big(m_\mu^2 s\big|F_1(m_\chi^2)-F_2(m_\chi^2)\big|^2+ \nonumber\\
&t~(|F_1(m_\chi^2)|m_\mu^2-|F_2(m_\chi^2)|s)\Big)+|q_L|^2 ~t\Big(\big|F_1(m_\chi^2)m_\mu^2-F_2(m_\chi^2)s\big|^2-\nonumber\\
&t~(|F_1(m_\chi^2)|^2m_\mu^2-|F_2(m_\chi)|^2s)\Big)\bigg)+t\leftrightarrow s~+\nonumber\\
 &\frac{2|q_e|^2(m_\mu^2-s-t)\left(m_\chi^2(\Gamma_\chi^2+m_\chi^2-s-t)+s~t\right)}{\left(\left(m_\chi^2-s\right)^2+m_\chi^2\Gamma_\chi^2\right)\left(\left(m_\chi^2-t\right)^2+m_\chi^2\Gamma_\chi^2\right)}\nonumber\\
 &\Big(m_\mu^2(s+t)(|F_1(m_\chi^2)|^2-F_1(m_\chi^2) F_2(m_\chi^2))+|F_2(m_\chi^2)|^2s~t\Big)\Bigg]\,,
 \label{eq:doubly-differential-muto3e-OLM}
 \end{align}
where $s\equiv(p_\mu-p_{e_1})^2$ and $t\equiv(p_\mu-p_{e_2})^2$, with kinematic limits given in  eq.~(\ref{eq_integration_limits}), and $\Gamma_\chi$ the total decay width of $\chi$. Similarly to Section \ref{sec:tree-level}, the dominant decay modes are $\chi\rightarrow e^-e^+,\overline{\nu_{L_1}}\nu_{L_1}, \overline{\nu_{L_2}}\nu_{L_2}$, with width:
\begin{align}
 \Gamma_\chi= \frac{g_\chi^2~m_\chi}{24\pi}\Big(|q_e|^2+3|q_L|^2\Big)\,.
\end{align}

\begin{figure}[t!]
\begin{center}
\includegraphics[width=0.49\linewidth]{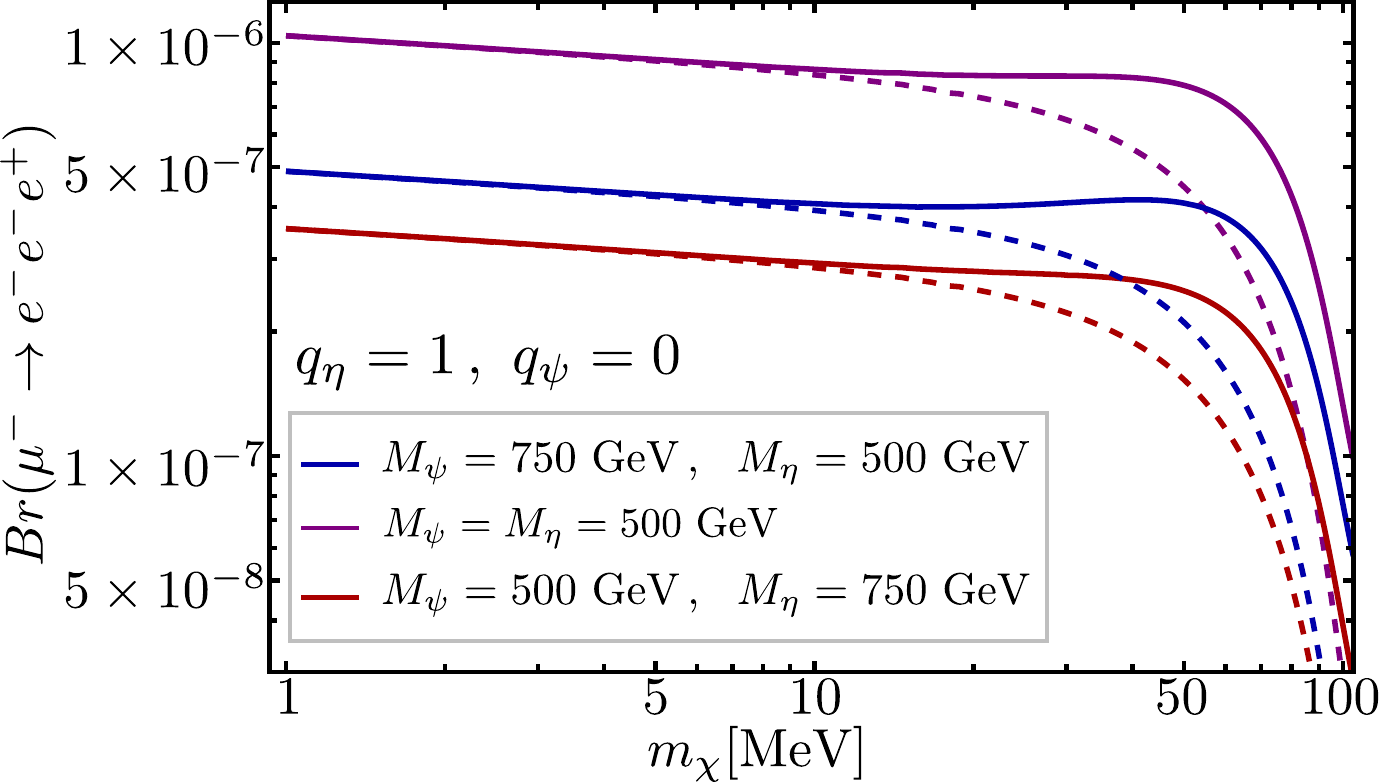}~~
\includegraphics[width=0.49\linewidth]{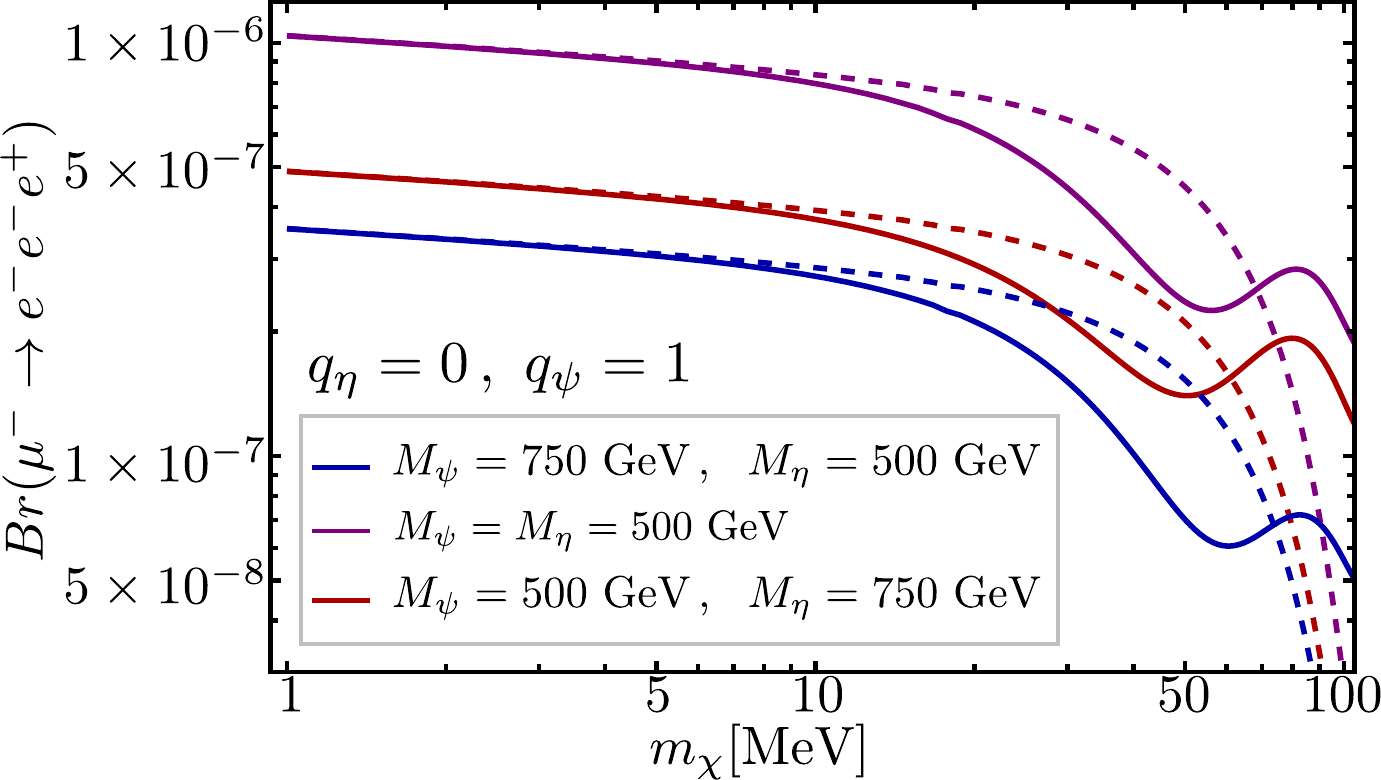}
\end{center}
\caption{Same as Fig.~\ref{fig:mu_to_echi_loop}, but for the process $\mu\rightarrow 3e$, assuming $q_L=1+q_e$.}
\label{fig:mu_to_3e_loop}
\end{figure}

\begin{figure}[t!]
\begin{center}
\includegraphics[width=0.49\linewidth]{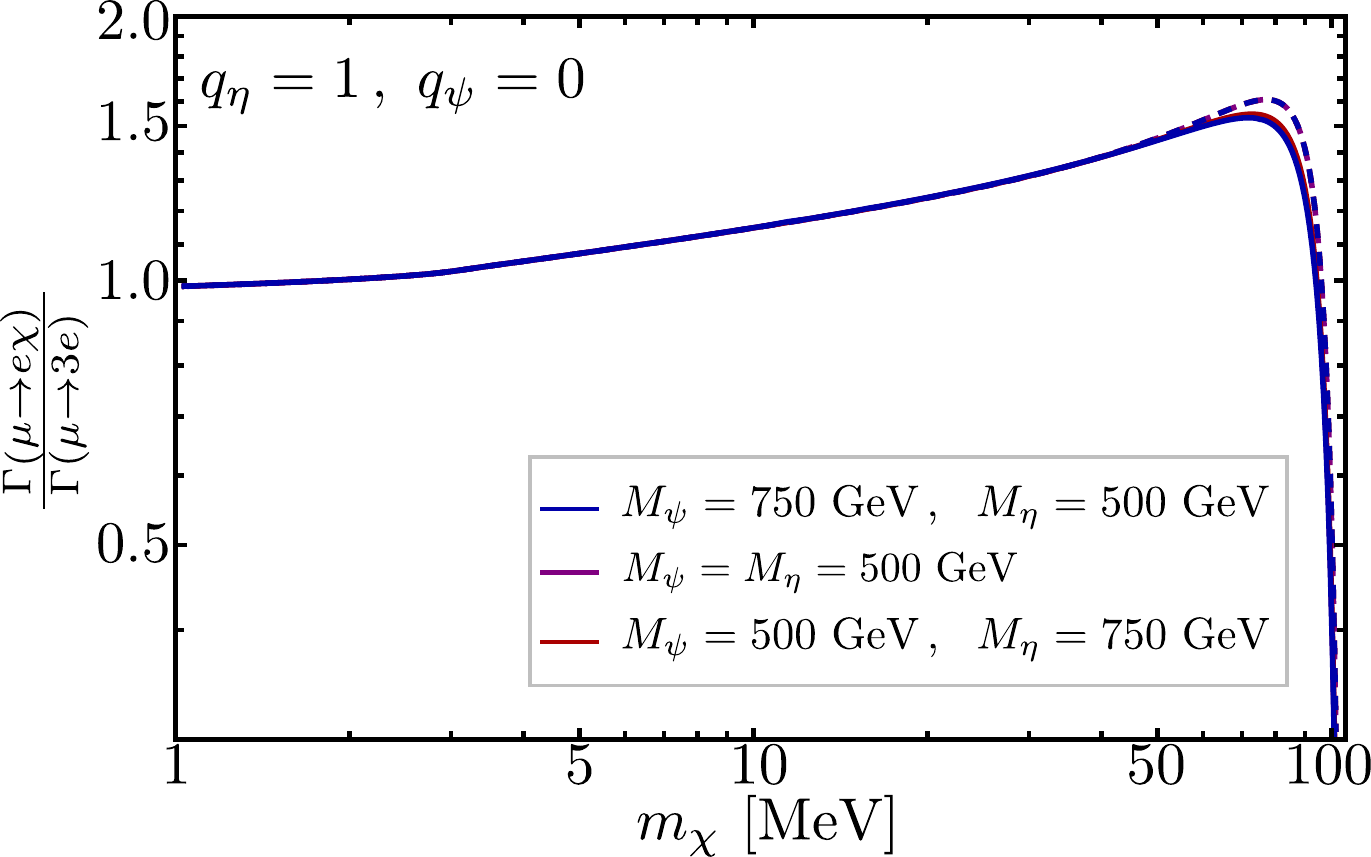}~~
\includegraphics[width=0.49\linewidth]{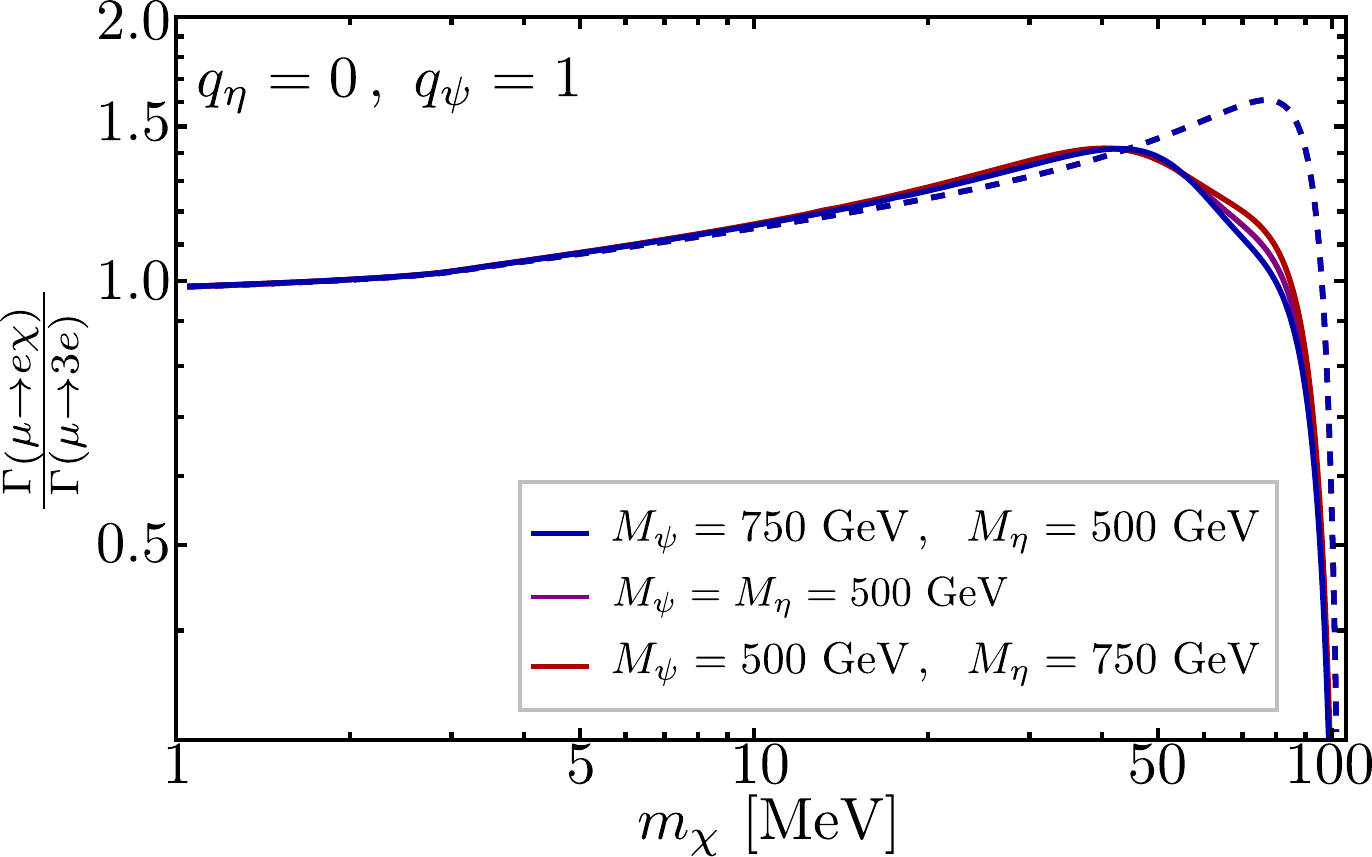}
\end{center}
\caption{Same as Fig.~\ref{fig:mu_to_echi_loop}, but for the ratio of rates $\Gamma(\mu\to e\chi)/\Gamma(\mu\rightarrow 3e)$, assuming $q_L=1+q_e$.}
\label{fig:ratio-of-rates_OLM}
\end{figure}

We show in Fig.~\ref{fig:mu_to_3e_loop},  the  branching ratio for $\mu\rightarrow 3e$ for the same choices of $q_\eta$ and $q_\psi$ as in Fig.~\ref{fig:mu_to_echi_loop},  and adopting $q_L=1+q_e$ (with $q_e=q_\psi+q_\eta)$, using the full result Eq.~(\ref{eq:doubly-differential-muto3e-OLM}) (solid lines) or setting the form factors $F_1=G_1=0$ (dashed lines). As for $\mu\rightarrow e\chi$, the form factors can be neglected when $m_\chi\ll m_\mu$, and only contribute to the rate when $m_\chi/m_\mu\gtrsim 0.1$. In Fig.~\ref{fig:ratio-of-rates_OLM} we show the ratio of rates  $\Gamma(\mu\to e\chi)/\Gamma(\mu\to 3e)$ as a function of $m_\chi$ for each of the cases. We find that the ratio is $\sim 1$. As in Section \ref{sec:tree-level}, this result can be understood analytically employing the narrow width approximation. Under this approximation, the decay rate for $\mu\to3e$ reads
\begin{align}
 \Gamma(\mu\to 3e)\simeq &\frac{ m_\mu }{4 \pi}\frac{|q_e|^2+|q_L|^2}{|q_e|^2+3|q_L|^2}\left(1-\frac{m_\chi^2}{m_\mu^2}\right)^2 \Bigg[\Big|F_1(m_\chi^2)\frac{ m_\mu}{ m_\chi}-F_2(m_\chi^2)\frac{ m_\chi}{m_\mu}\Big|^2+2|F_1(m_\chi^2)-F_2(m_\chi^2)|^2\Bigg]+\nonumber\\
 &\frac{g_\chi^2|q_e|^2}{16\pi}\frac{m_\chi^2}{m_\mu}\left(1-2\frac{m_\chi^2}{m_\mu^2}\right)\left(2\left(|F_1(m_\chi^2)|^2-F_1(m_\chi^2)F_2(m_\chi^2)\right)+|F_2(m_\chi^2)|^2\frac{m_\chi^2}{m_\mu^2}\right)\,.
 \label{eq:Rate-3e-NWA-OLM}
 \end{align}
Also in this scenario one finds an apparent divergence when $m_\chi\rightarrow 0$, however the factor $1/m_\chi$ in the rate is cancelled by the factor $m_\chi$ implicitly contained in the form factor $F_1$. As a result, in the limit $m_\chi\rightarrow 0$ the rate for $\mu\to 3e$ is finite and comparable to the rate for $\mu\rightarrow e\chi$, quite independently of the masses and charges of the particles in the loop.

Given that the current limits on the processes $\mu\to 3e$ and $\mu\to e\chi$, we expect the former to yield the strongest limits on this scenario. This is apparent from Fig.~\ref{fig:mu_to_3e_loop}: the three choices of parameters are allowed by the current constraints on $\mu\to e\chi$, but several orders of magnitude above the SINDRUM limit ${\rm Br}(\mu\to 3e)< 1.0 \times 10^{-12}$.

\section{Conclusions}\label{sec:conclusions}

We have studied in detail the lepton flavor violating process $\mu\rightarrow e \chi$, with $\chi$ a massive gauge boson arising from the spontaneous breaking of a local $U(1)$ symmetry. We have constructed the most general effective interaction between a muon, an electron and a massive gauge boson, and we have calculated the decay rate in terms of the corresponding form factors. The decay rate presents terms inversely proportional to the inverse of the $\chi$-boson mass, corresponding to the decay into the longitudinal component of the $\chi$-boson, which naively lead to an enhancement of the rate when $\chi$ is very light. 

We have constructed two gauge invariant and renormalizable models where the decay $\mu\rightarrow e\chi$ is generated either at tree level or at one loop. We have analyzed the behavior of the rate in the limit  $m_\chi\ll m_\mu$, and we have explicitly checked that the rate remains finite. We have also calculated the expected rate for the process $\mu\to 3e$, mediated by an off-shell $\chi$. For these two models, the ratio of rates of $\mu\to e\chi$ and $\mu\to 3e$ is ${\cal O}(1)$ in the range of $\chi$-masses considered. Correspondingly, and in view of the current limits on $\mu\to 3e$ from the SINDRUM collaboration, it would be necessary an improvement of experiments searching for $\mu\to e\chi$ of at least 5-6 orders of magnitude compared to the TWIST sensitivity in order to observe a signal.

\section*{Acknowledgements}

This work has been supported by the Collaborative Research Center SFB1258, by the Deutsche Forschungsgemeinschaft (DFG, German Research Foundation) under Germany's Excellence Strategy - EXC-2094 - 390783311, by the projects CB-250628 (Conacyt), Fondo SEP-Cinvestav number 142 and Cátedra Marcos Moshinsky (Fundación Marcos Moshinsky), as well as by the scholarships for Marcela Marín’s Ms. Sc. and Ph. D. Theses (Conacyt). P. R. is indebted to Denis Epifanov for suggesting this fascinating research topic. We have benefited from  enriching discussions with our experimental colleagues Eduard de la Cruz Burelo and Iván Heredia de la Cruz. We especially thank  Xabi Marcano and Julian Heeck for insightful remarks on the manuscript.

\bibliographystyle{JHEP-mod}
\bibliography{references}

\end{document}